\documentclass[conference]{IEEEtran}
\IEEEoverridecommandlockouts
\usepackage{cite}
\usepackage{amsmath,amssymb,amsfonts}
\usepackage{algorithmic}
\usepackage{graphicx}
\usepackage{url}
\usepackage{tabularx}
\usepackage{ragged2e}
\usepackage{textcomp}
\usepackage{xcolor}
\usepackage[caption=false]{subfig}
\graphicspath{{./figures/}}
\def\BibTeX{{\rm B\kern-.05em{\sc i\kern-.025em b}\kern-.08em
    T\kern-.1667em\lower.7ex\hbox{E}\kern-.125emX}}
\begin{document}

\title{Using AI for Mitigating the Impact of Network Delay in Cloud-based Intelligent Traffic Signal Control}

% \author{\IEEEauthorblockN{1\textsuperscript{st} Rusheng Zhang}
% \IEEEauthorblockA{\textit{Department of Electrical and Computer Engineering} \\
% \textit{Carnegie Mellon University}\\
% Pittsburgh, USA \\
% rushengz@andrew.cmu.edu}
% \and
% \IEEEauthorblockN{2\textsuperscript{nd} Xinze Zhou}
% \IEEEauthorblockA{\textit{Department of Electrical and Computer Engineering} \\
% \textit{Carnegie Mellon University}\\
% Pittsburgh, USA \\
% rushengz@andrew.cmu.edu}
% \and
% \IEEEauthorblockN{3\textsuperscript{rd} Ozan K. Tonguz}
% \IEEEauthorblockA{\textit{Department of Electrical and Computer Engineering} \\
% \textit{Carnegie Mellon University}\\
% Pittsburgh, USA \\
% rushengz@andrew.cmu.edu}
% % \and
% % \IEEEauthorblockN{4\textsuperscript{th} Given Name Surname}
% % \IEEEauthorblockA{\textit{dept. name of organization (of Aff.)} \\
% % \textit{name of organization (of Aff.)}\\
% % City, Country \\
% % email address or ORCID}
% % \and
% % \IEEEauthorblockN{5\textsuperscript{th} Given Name Surname}
% % \IEEEauthorblockA{\textit{dept. name of organization (of Aff.)} \\
% % \textit{name of organization (of Aff.)}\\
% % City, Country \\
% % email address or ORCID}
% % \and
% % \IEEEauthorblockN{6\textsuperscript{th} Given Name Surname}
% % \IEEEauthorblockA{\textit{dept. name of organization (of Aff.)} \\
% % \textit{name of organization (of Aff.)}\\
% % City, Country \\
% % email address or ORCID}
% }
\author{\IEEEauthorblockN{Rusheng Zhang\IEEEauthorrefmark{1},
Xinze Zhou\IEEEauthorrefmark{2}, Ozan K. Tonguz\IEEEauthorrefmark{3}}
\IEEEauthorblockA{Department of Electrical and Computer Engineering, Carnegie Mellon University\\
PA 15213-3890, Pittsburgh, USA\\
Email: \IEEEauthorrefmark{1}rushengz@andrew.cmu.edu,
\IEEEauthorrefmark{2}xinzez@andrew.cmu.edu,
\IEEEauthorrefmark{3}tonguz@ece.cmu.edu}}
\maketitle

\begin{abstract}
The recent advances in cloud services, Internet of Things (IoT)  and  Cellular networks have made cloud computing an attractive option for intelligent traffic signal control (ITSC). Using cloud computing significantly reduces the cost of cables, installation, number of devices used, and maintenance.  ITSC systems based on cloud computing lower the cost of the ITSC systems and make it possible to scale the system by utilizing the existing powerful cloud platforms such as Google Cloud Platform (GCP), Microsoft Azure, and Amazon Web Service (AWS).

While such systems have significant potential, one of the critical problems that should be addressed is the network delay. It is well-known that network delay in message propagation is hard to prevent, which could potentially degrade the performance of a system or even create safety issues for vehicles at intersections.

In this paper, we introduce a new traffic signal control algorithm based on reinforcement learning, which performs well even under severe network delay. The framework introduced in this paper can be helpful for all agent-based systems using remote computing resources where network delay could be a critical concern. Extensive simulation results obtained for different scenarios show the viability of the designed algorithm for coping with network delay.
\end{abstract}


\section{Introduction}
The recent advances in technologies such as the Internet of Things (IoT), cloud services, mobile computing, and  cellular networks have made cloud computing an attractive option for intelligent traffic signal control (ITSC). Such an approach can significantly reduce the cost of cables, installation, devices used, and maintenance. Therefore, several cloud computing based traffic systems were proposed recently \cite{prasad2011adaptive, hahanov2013cloud, hahanov2014cyber}.  

To reduce cost, in most of the proposed systems, the vehicles and traffic signal control agents connect to the cloud in a wireless manner, e.g., through LTE, LTE-A, or 5G connections. While such systems have a huge potential, one of the critical problems that needs to be addressed is the network delay. Unexpected  network delay could potentially degrade the performance of the system, or even create serious safety issues. 

Cloud computing is also a very interesting option for Virtual Traffic Lights (VTL) \cite{ferreira2010self, tonguzred, zhang2018virtual} technology, which proposes to equip traffic light devices in vehicles and yield infrastructure-free traffic signal control. While initially proposed to implement with Vehicle-to-vehicle (V2V) communications, VTL could take advantage of the rapid developments in cloud computing, smart phone applications as well as cellular networks (such as LTE, 5G), and can be implemented with a cloud and smart phone based system. Clearly, in this case, network delay is a critical concern because even if one of the devices can't receive traffic phase information in time, accidents might happen.

It is well-known that unexpected network delay due to random events during propagation is a very hard problem to avoid entirely. An ITSC algorithm which tolerates a certain degree of delay is therefore a more realistic solution. However, implementing this approach is non-trivial since there isn't a comprehensive ITSC model that takes network delay into consideration. Therefore, Deep Reinforcement Learning (DRL), an artificial intelligence (AI) algorithm that has recently become very popular \cite{mnih2013playing} (observe that DRL can find optimal solutions for complicated problems without a comprehensive analysis of the model itself), could be a good option to explore for the aforementioned problem. 

In this paper, we design an ITSC algorithm that takes network delay into consideration. The algorithm is based on DRL and is able to perform well under a certain level of network delay. The algorithm could be helpful for cloud-based systems where network delay is a concern. By using the algorithm proposed in this paper, these systems can be implemented in the real-world.

\section{Related Works}
Cloud computing has been widely explored for vehicular usage in recent years. To name a few, \cite{gerla2012vehicular, lee2014vehicular} discusses vehicular cloud in terms of the design principle, architecture, and research issues pertaining to the vehicular cloud. \cite{al2015intelligent}, on the other hand, proposes a traffic monitoring system based on cloud computing and IoT that monitors vehicles using RFID and wireless sensor networks.  \cite{feng2017hvc} proposes a hybrid cloud computing framework of Dedicated Short-Range Communication (DSRC) and cellular network to utilize resources of OnBoard Unit (OBU), RoadSide Unit (RSU) as well as centralized cloud to reduce cellular network usage and increase the processing speed. 

The latency of cloud computing for vehicular use is a well known issue and many different approaches have been proposed thus far to address this concern. As an example, \cite{balasubramanian2020low} introduced an infrastructure model that forms low latency vehicular service clouds on-the-fly, and reduces latency by using edge computing. Several other studies introduce fog computing into vehicular environments to enable low-latency services \cite{sookhak2017fog, xue2018fog}. Unfortunately, all these solutions require additional infrastructure, hence are not suitable for the ITSC systems considered in this paper.

% \cite{jaworski2011cloud} explores the concept of cloud-computing Intelligent transportation System (ITS) and report the development of a prototype.   \cite{ma2012user}  
Several other research studies aim to use vehicular cloud for traffic signal control. \cite{prasad2011adaptive} propose a cloud based traffic signal control system based on camera detection. \cite{hahanov2013cloud} introduces another cloud based system that gathers data through RFID from vehicles and generates 'green wave traffic'. These studies focus on utilization of the computation power in the cloud and assume ideal network connections; hence, network latency could potentially degrade the performance of these systems. \cite{hahanov2014cyber}  proposes cyber-physical system that is based on virtual cloud and cellphones. The author makes an interesting observation that the physical infrastructures will gradually move to the virtual space, substituted by display units inside vehicles as well as virtual cloud, which is the main idea of Virtual Traffic Lights (VTL).

VTL is a technology initially proposed based on DSRC technology to yield fully infrastructure-free traffic signal system displayed in vehicles \cite{ferreira2010self, tonguzred, zhang2018virtual}. While the initial concept is based on DSRC, clearly, such idea can also be implemented using smart phones and cloud (refer to Section \ref{ss:PS} and Figure \ref{C-VTL}) to utilize the powerful cloud platforms and services. One of the key technical issues in migrating from a DSRC based distributed method to a centralized cloud-based method is the network latency, as vehicles need to forward real-time information to the cloud server, and display the in-time traffic signal command received from the cloud.

On the other hand, several research groups have also explored the use of  DRL in ITSC recently \cite{genders2016using, gao2017adaptive, mousavi2017traffic, wei2018intellilight}. While these studies show that DRL algorithm indeed optimizes the ITSC systems, DRL only outperforms existing traffic optimization methods in a minor way (if any). A more interesting applicatio of DRL in traffic signal control is to apply it over partially detected ITSC (PD-ITSC) where only a proportion of vehicles are detected \cite{zhang2018partially, zhang2019partial}. Since there isn't any existing analytic model for such systems, DRL can serve as a powerful  tool that can solve the problem quickly. The success of applying DRL to the aforementioned systems indicates that DRL could be a highly viable method for  traffic signal control where network delay could be a problem.

In this paper, we introduce a DRL based algorithm to address the concern on network delay. While the algorithm is primarily designed for migrating the VTL system from distributed DSRC system to cloud-based VTL (C-VTL) system, the results and findings in this paper suggest that the framework introduced could be potentially useful for all the aforementioned ITSC systems to build a delay-tolerant traffic signal control algorithm (formally introduced in Section \ref{ss:delay-tolerant}).

\section{Problem Statement}
\label{ss:PS}
\subsection{Cloud-based ITSC System}
\label{ss:system}
For convenience, we first introduce two illustartive examples of cloud-based ITSC systems. Note that the algorithm is also useful for other systems as well; the systems introduced below are just two common cases of the many applicable examples.

\subsubsection{Cloud Traffic System (CTS)}
\label{SSS:CTS}
\begin{figure} [ht]
    \centering
  \subfloat[Conceptual figure showing the Cloud traffic system (CTS) \label{CTS}]{%
       \includegraphics[width=0.8\linewidth]{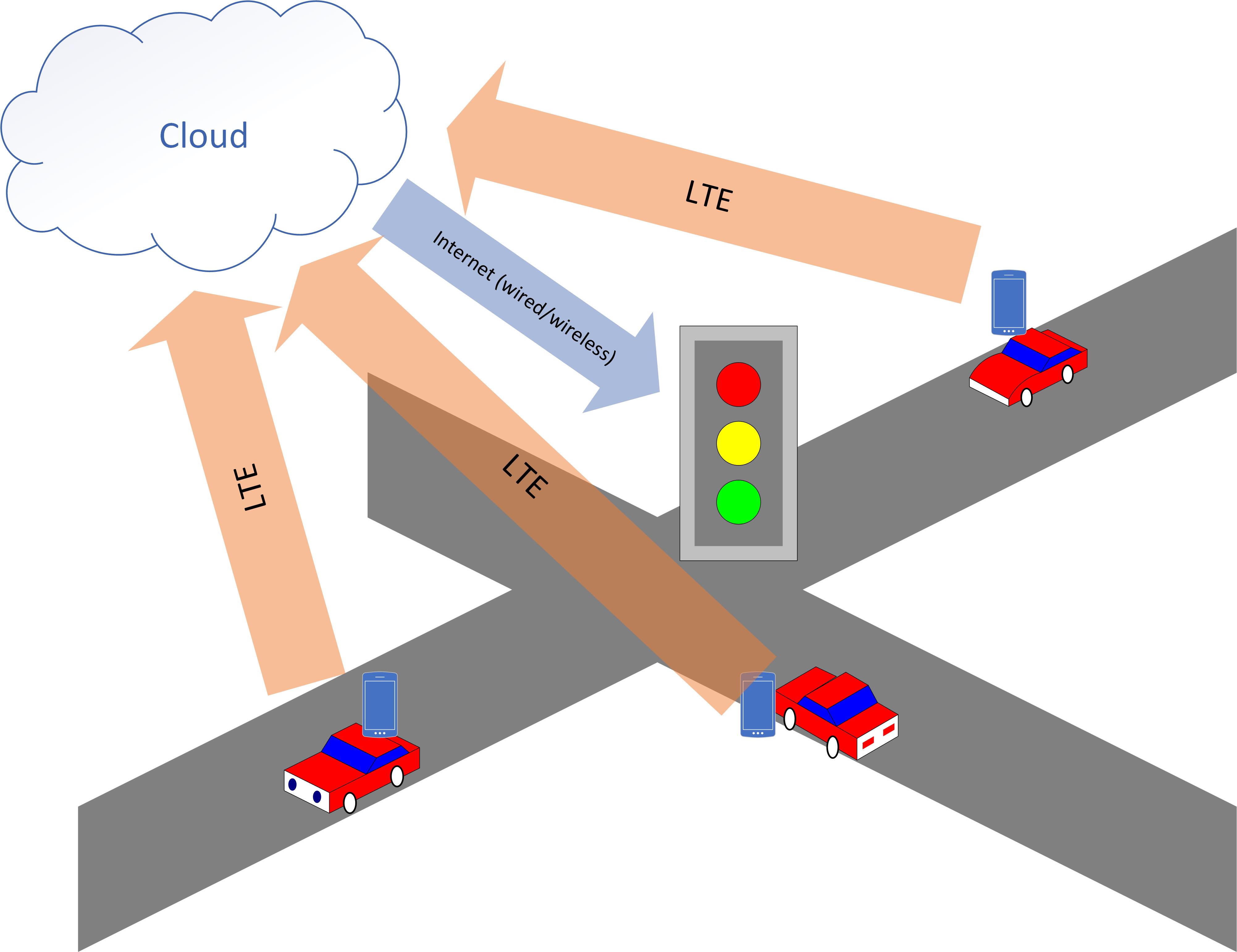}}
       %\hfill
    \\
  \subfloat[Conceptual figure of the Cloud based Virtual Traffic Lights (C-VTL) system \label{C-VTL}]{%
        \includegraphics[width=0.8\linewidth]{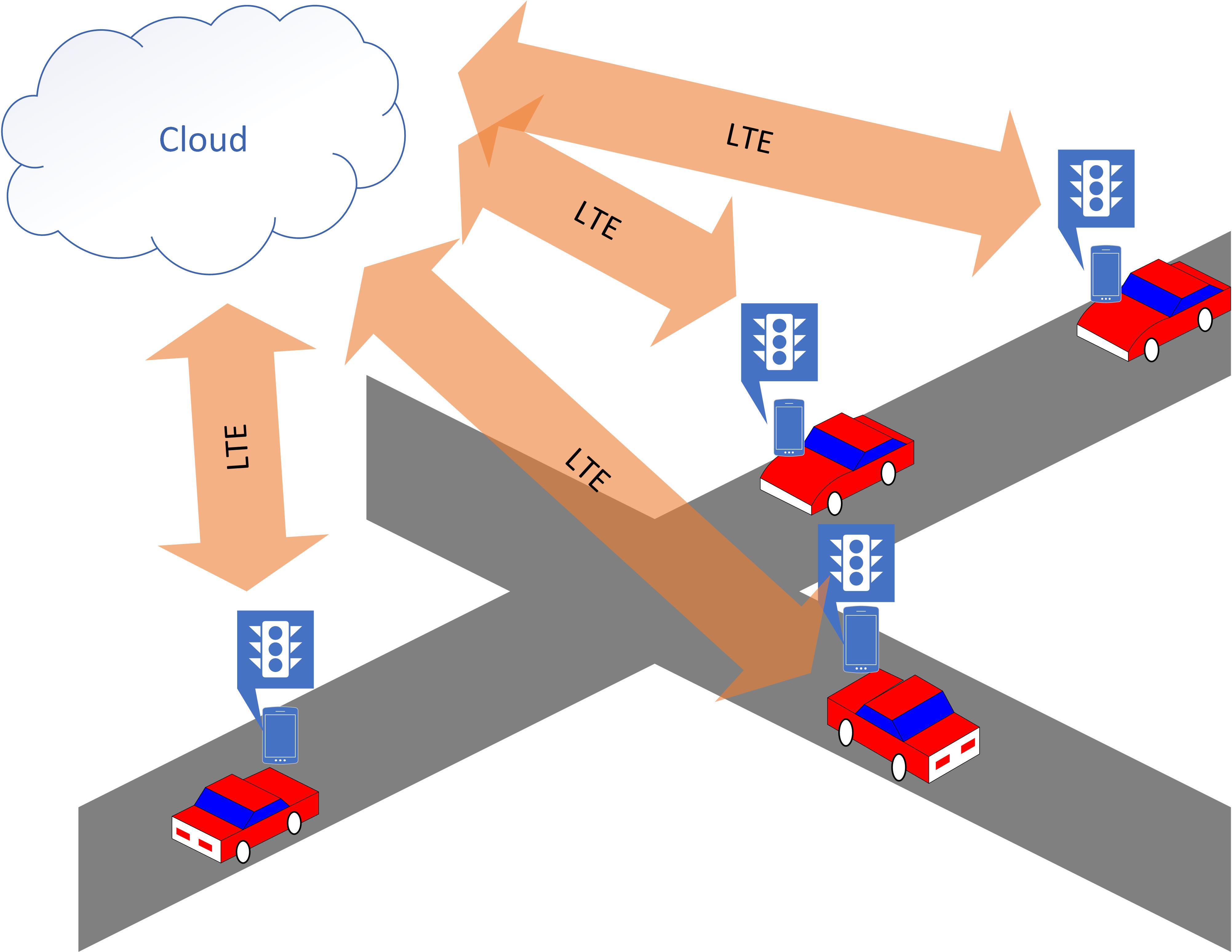}}
  \caption{Two possible cloud based intelligent traffic light system of interest}
  \label{fig1} 
\end{figure}

Figure \ref{CTS} shows the Cloud Traffic System (CTS) of interest. In this ITSC system, vehicles forward their geo-information to the cloud server through a cellular network (e.g., LTE) using cellphone or other devices in a periodical manner. The cloud server gathers the traffic information sent from vehicles and decides the traffic signal phase(s) at the intersection. The cloud server then forwards the decision to traffic light(s) controller. The controller receives the command and keeps or switches the traffic signal phase accordingly.

Obviously, if there is a latency between vehicles and the cloud server, the cloud server won't be able to make correct decision in  a timely manner, and traffic phase command will not be able to arrive in time. Meanwhile, the network delay for different vehicles  can be different, if the cloud server is not aware of this fact,  the server might make incorrect decisions that could lead to undesirable consequences.

\subsubsection{Cloud-based Virtual Traffic Lights (C-VTL)}
Cloud-based Virtual Traffic Lights (C-VTL) system is another cloud-based system that has huge potential. This system could be used in future generations of traffic lights. The concept of the system is shown in figure \ref{C-VTL}, each vehicle  has a cellular device with a display unit (e.g., a smartphone). The geo-information of each vehicle is forwarded to the cloud server via a cellular network (e.g., LTE) for decision making. The cloud server receives the messages and make traffic light phase decision for the corresponding intersection accordingly, and send the decision to all relevant vehicles. Each vehicle receives traffic light phase decision and display the traffic light phase (i.e., green light, red light or amber light) for itself on its display unit.

Network latency can be even more critical for C-VTL system in comparison to the CTS system. The network delay issue discussed in Section \ref{SSS:CTS} is also a concern for C-VTL systems. Furthermore, for C-VTL, since each vehicle displays the traffic signal information to its driver on its own display unit (as opposed to CTS condition, where all vehicles look at the same traffic light), with different network delay, the traffic light phase on different device might not even be synchronized. Such circumstances can even result in traffic accidents if two vehicles in orthogonal approaches both have green lights due to the aforementioned network delay.

From the discussion above, one can observe that network delay is a critical concern for cloud-based ITSC systems, both CTS and C-VTL. In the next subsection, we formally introduce the network delay problem and the technical task that this paper aims to solve.

\subsection{Delay-tolerant algorithm}
\label{ss:delay-tolerant}

As it is very hard (if not impossible) to completely eliminate network delay between mobile devices and cloud server, in this paper, we consider network delay as an inevitable event or a given design parameter that should be taken into account from the beginning. For the traffic signal control problem, we categorize the network delay as \textit{forward network delay} and \textit{backward network delay}. The delay of vehicles forwarding the geo-information to the server is denoted as \textit{forward network delay} or \textit{forward delay}; the delay of the server  forwarding traffic light phase command to the device (traffic light in CTS or mobile device in C-VTL) is denoted as \textit{backward network delay} or \textit{backward delay}. For both kinds of delay, we assume a maximum value: $t_{d1}$ for maximum forward delay and $t_{d2}$ for maximum backward delay. Note that $t_{d1}$ and $t_{d2}$ are the design parameters that are known when designing the algorithm for a certain intersection.

 As the main goal of this paper, with the known the maximum backward delay $t_{d1}$ and maximum forward delay $t_{d2}$, we want to find the optimized traffic signal control algorithm, we denote this algorithm as a \textit{delay-tolerant algorithm}.

\section{Methodology}
\subsection{Overview}
\label{S:overview}
As stated above, in this paper, we consider the network delay as a given or an inevitable event that we cannot control, and design the algorithm with known maximum forward delay $t_{d1}$ and backward delay $t_{d2}$. The two parameters can be considered as design parameters for the algorithm for realistic applications. One can increase the two parameters to make the algorithm 'safer', but as a trade-off, the performance of the algorithm will degrade.

We consider vehicles forwarding messages to the server periodically and the server periodically issues the traffic light phase command and sends it back with the same period. For convenience, we discretize time by the period as time steps. In the rest of the paper, the unit of time, such as $t_{d1}$, $t_{d2}$ are all considered to be steps.

The fundamental idea is: for the cloud server, instead of doing decision for current time, it makes the decisions for a short time period length $t_{d2}$ \textbf{in the future}. In other words, assume current  time step to be $t_c$, the server makes decisions for the whole period that starts at $t_c$ and ends at $t_c+t_{d2}$. In this way, since the maximum backward delay is $t_{d2}$, the message is guaranteed to be able to arrive before $t_c+t_{d2}$. Analogously, instead of using the messages from vehicles at time steps $t_c$ for decision making, we use the information at time $t_c-t_{d1}$. Since the maximum forward delay is $t_{d1}$, all vehicles' messages at time $t_c-t_{d1}$ will be received at time $t_c$.

To that end, each message forwarded by the vehicle should
be associated with a timestamp, say $t_1$, and the message will be buffered by the server in its memory. The server retrieves all the messages with timestamp $t_1$ at the time $t_1+t_{d1}$ and process it with a DRL algorithm (discussed in Section \ref{SS:DQL}). Each action output from the DRL agent will also be associated with a timestamp, indicating the action is made for the time specified on the timestamp. For the action made at time $t_c$, the timestamp for that action should be $t_c+t_{d2}$.
\begin{figure}[ht]
    \centering
    \includegraphics[width=.9\linewidth]{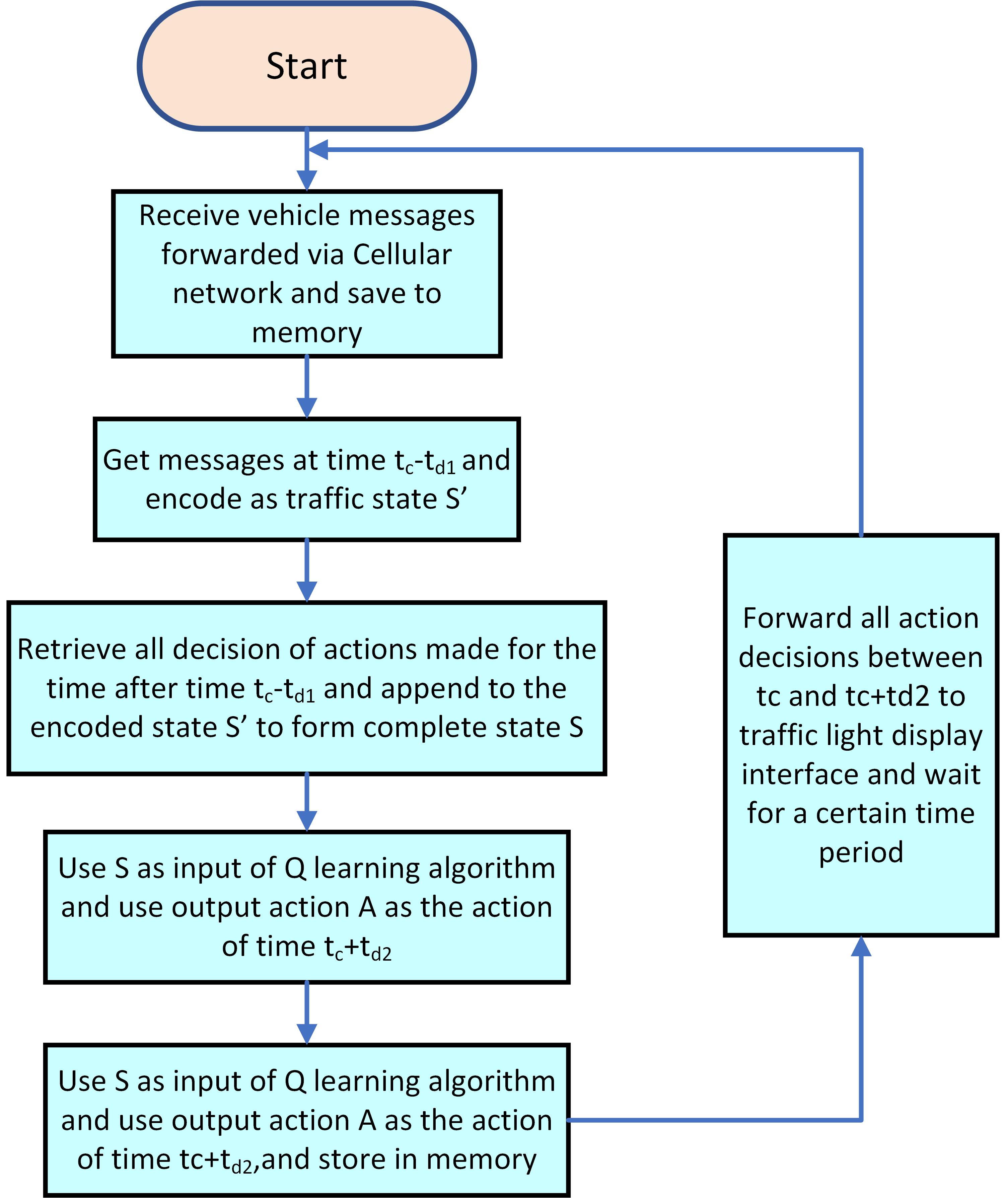}
    \caption{A diagram of the algorithm}
    \label{fig:flowchart}
\end{figure}

Figure \ref{fig:flowchart} concludes the algorithm on the server side. Periodically, it collects messages forwarded by the vehicles and saves them to the memory. At each time step $t_c$, the server retrieves all messages with timestamp $t_c-t_{d1}$, as well as all the previous decisions made for the time step after $t_c-t_{d1}$ to form the observation vector. The reason that we need to append these actions to the state representation of the DRL algorithm is the following: these actions are the decisions made by the server in previous steps and already forwarded to the traffic signal display unit. To preserve consistence, these actions can not be modified. Since the DRL agent does not have memory, the DRL agent needs to know what actions have previously decided to make accurate decisions. The DRL algorithm will output the action for time step $t_c+t_{d2}$, and such action, together with all the actions made for time after $t_c$ will be sent to the traffic light display unit (traffic light for CTS or the display unit in each vehicle for C-VTL).

In the next subsection, we will elaborate on the DRL algorithm.

\subsection{Deep Reinforcement Learning}
\label{SS:DQL}
Section \ref{S:overview} shows the overall process of the algorithm. Within this process, the DRL algorithm is treated as a black box that inputs the observation at $t_c-t_{d1}$ and decides actions after time step $t_c-t_{d1}$ and outputs the action for $t_c+t_{d2}$. In this subsection, we will give more details on the DRL algorithm.

DRL is an algorithm that trains an agent which interacts with environment to correctly select actions which optimize a certain goal. There are different DRL algorithms, in this paper, we choose Deep Q learning (DQL) algorithm \cite{mnih2013playing}. In Q learning algorithm, the agent learns the action-value function known as 'Q-value', $Q(S_t, A_t)$. Q-value is a function that maps current state $S_t$ to the expected cumulative discounted future reward given state and action. At every time step, the agent updates its Q value with the following fomula:
\[
\begin{split}
    Q(S_t,A_t) = Q(S_t,A_t) + \alpha(R_{t+1}+\gamma \max{Q(S_{t+1},A_t)}\\-Q(S_t,A_t))
    \end{split}
\]
In the equation, $\alpha$ is the learning rate of the algorithm and $R_t$ is the reward for time $t$. To adopt the algorithm for a specific problem, one needs to define the observation state, action, and reward. In this paper, we use similar state observation, action and reward defined in \cite{zhang2018partially} with minor modifications.

\subsubsection{State observation}
 For traffic state representation, we choose to use Compact State Representation (CSR)  proposed in \cite{zhang2018partially}, as CSR representation was proved to be effective in previous work. In this paper, the state representation for the DRL is the CSR state representation appending the previously decided action from $t_c-t_{d1} $ to $t_c+t_{d2}$. The detail of the state representation is shown in Table \ref{tab:stateRep}.

\begin{table}[ht]
    \centering
    \caption{state representation}
    \label{tab:stateRep}
\begin{tabularx}{.95\linewidth}{|p{2cm}||X|}
\hline
\textbf{Information}&\textbf{Representation}\\
\hline
\RaggedRight
Detected car count  &Number of detected vehicles in each approach (normalized by maximum capacity of the lane)\\
\hline
Distance & Distance to nearest detected vehicle on each approach\\
\hline
Phase time  & How much time elapsed in current phase (in seconds)\\
\hline
Amber phase  & Indicator of amber phase; 1 if currently in amber phase, otherwise 0 \\
\hline
Current phase & An integer to represent current traffic signal phase\\
\hline

Current time (optional) & Current time of the day\\
\hline
Previously decided actions & Previously decided actions for time step $t_c-t_{d1}$ to $t_c+t_{d2}$\\
\hline
\end{tabularx}
\end{table}

\subsubsection{Action}
The relevant actions for the a traffic display unit is either to keep the current  phase or to switch to the next one. The action is encoded to be either 1 or 0, where keeping the current phase is 1 and switching to the next phase is 0.

\subsubsection{Reward}
We use the same reward in \cite{zhang2018partially} as it is proven to be effective:
\begin{equation}
R_t = -\sum_{c\in C} \frac{1}{v_{\max, c}}[v_{\max,c}-v_{c}(t)]
\label{eq:reward}
\end{equation}
In the equation, $R_t$ denotes reward at time $t$, $v_{\max, c}$ is the maximum speed of vehicle $c$, $v_{c}(t)$ is the vehicle $c$'s actual speed at time $t$ and $C$ is the set of all vehicles of interest.
\section{Simulations}
\label{S:simulations}
\subsection{Simulation settings}
We use Gym Trafficlights \cite{gym_trafficlight}, a traffic environment developed based on SUMO \cite{SUMO2018} and OpenAI Gym \cite{brockman2016openai} for reinforcement learning related studies. The simulation code can be found in \cite{cloud_dqn}. We tested the algorithm on the \textit{simple\_environment} provided by Gym Trafficlights with 3 different but realistic regimes of car flow :
\begin{enumerate}

\item \textit{Sparse}: Only very few vehicles come to the intersection, which corresponds to a midnight situation, e.g., the car flow at 2 AM in the morning.
    \item \textit{Dense}: Many vehicles come to the intersection, which corresponds to the rush hour situation, e.g., car flow at 8 AM.
    \item \textit{Medium}: Intermediate car flow, which corresponds to regular hour traffic, e.g., car flow at 2 PM.
\end{enumerate}

Due to randomness, not every training will yield optimal result (this lack of robustness might be alleviated by adopting an LSTM strategy and perform a curriculum learning, see \ref{ss:discussion}). Therefore, for each scenario, we conduct three independent training processes and choose the best-performing network. 

The performance of the algorithm is evaluated by the average waiting time of vehicles at the intersection, using 5 independent 1-hour simulations. Note that even though the algorithm has two design parameters $t_{d1}$ and $t_{d2}$, the performance of the algorithm is only affected by the sum of the two. Therefore, we evaluate the performance over the total delay, namely, $t_{d1} + t_{d2}$.

\subsection{Results}
\label{ss:result}
\begin{figure} [htb]
    \centering
  \subfloat[Performance over sparse car flow \label{fig:sparse}]{%
       \includegraphics[width=0.9\linewidth]{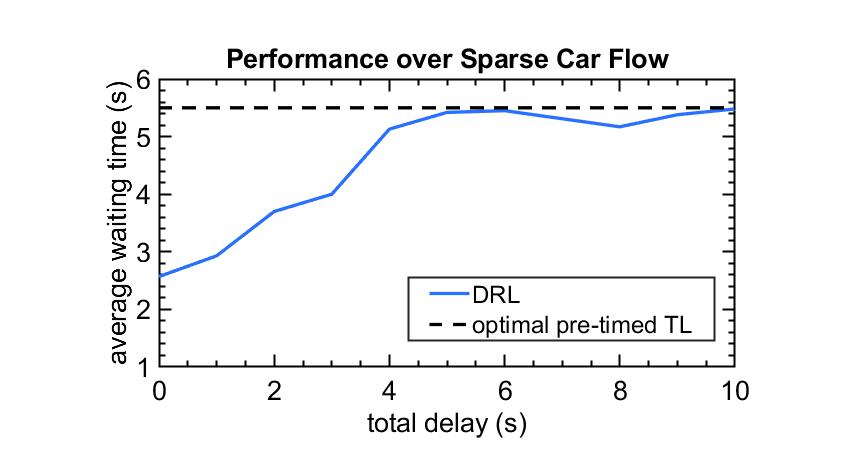}}
       %\hfill
    \\
  \subfloat[Performance over medium car flow \label{fig:medium}]{%
        \includegraphics[width=0.9\linewidth]{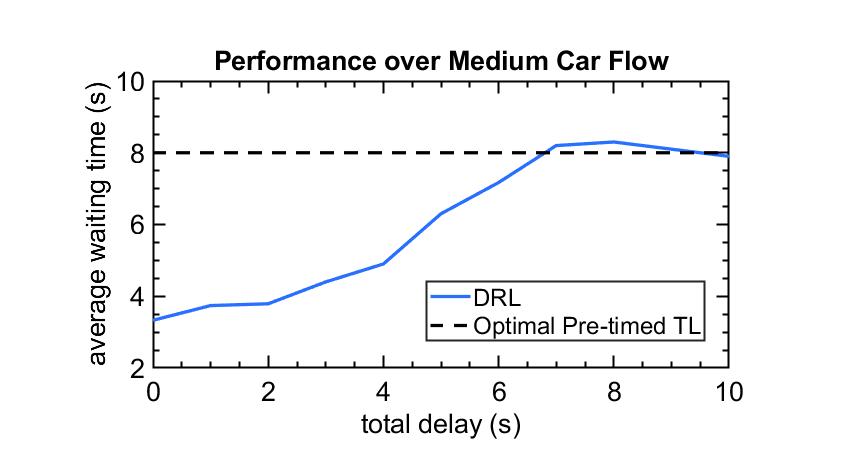}}
        \\
  \subfloat[Performance over dense car flow \label{fig:dense}]{%
        \includegraphics[width=0.9\linewidth]{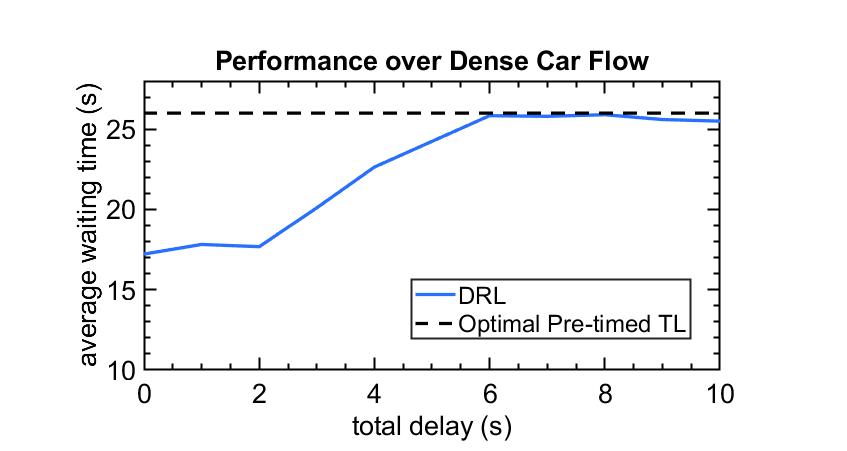}}
  \caption{Performance over different car flow}
  \label{fig:performance} 
\end{figure}

Figure \ref{fig:performance} shows the performance of the algorithm for different car flows. For all three cases, we find a common trend: the performance curve can be divided into two phases. In phase 1,  the average waiting time gradually increases with the total maximum delay.  The convexity switches from convex upward to convex downward during this phase. When the delay is larger than a certain value, the average waiting time will stop increasing and the curve enters phase 2. In phase 2, the average waiting time does NOT increase with maximum total delay anymore. Instead, it stays constant at the performance of an optimal pre-timed traffic light.  This two-phase behavior or trend agrees with our intuition: when the delay is very large, the detection of individual vehicles is less useful; hence, the optimized solution is just the optimal pre-timed traffic light (we also have confirmed this visually in the simulation that under when the total delay is high, the traffic light collapses to a fixed-time traffic light with optimal phase-split and period). 

It is also interesting to observe that, in all the three cases, the experienced performance loss is relatively small when delay goes up to 2 seconds. This behavior suggests that the algorithm will be able to tolerate a certain degree of delay without a major performance loss. Even with high delay, the algorithm will still converge to the optimal pre-timed signal with the best phase split and period for that intersection. This will out-perform a fixed traffic light that's not optimal to the current car flow, which is very common at most of today's intersection. For example, if we consider a fixed-time traffic light of 30 seconds phase time (which is not the optimum, but very common), it will yield 13 seconds of waiting time under sparse car flow, 15 seconds of waiting time under medium car flow and 27 seconds of waiting time under dense car flow. Observe that these values are much higher than the performance shown in figure \ref{fig:performance}.

For the sparse flow, we observe that the period of phase 1 is roughly 4 seconds, which is shorter than medium and dense car flow. This is due to the fact that when the car flow is sparse, it is more important to detect each vehicle's current location to optimize traffic signal. While the trends are the same, we observe that the percentage  performance loss with delay is much smaller for the dense flow than the sparse and medium flow: it is 50\% for small and medium flow but less than 30\% for dense flow. The reason behind this is the following: when there are more vehicles, the information of vehicles' current position is not as important, and the optimization of traffic is therefore less effective. The fact that vehicles' position helps optimizing traffic in a smaller way when car flow is dense indicates that the algorithm can tolerate higher delay in dense traffic condition and is very robust under heavy traffic flows. When the traffic flow is high, the delay is expected to be higher because of the limited bandwidth of cellular network; hence, it's a desirable feature that the algorithm will be able to tolerate higher delay in dense traffic condition.

\section{Discussion}
\label{ss:discussion}
Based on the simulation results shown in figure \ref{fig:performance}, one can make a recommendation on how to choose the design parameters $t_{d1}$ and $t_{d2}$. Since in phase 1 (defined in section \ref{ss:result}), the convexity switches from convex upward to convex downward, this means that before the convexity switching point, the waiting time is relatively flat than after the switching point. Therefore, we can choose the convexity switching point to be our total maximum delay to achieve a balance or trade-off between performance and safety. Namely, 3 seconds for intersection with sparse car flow, 4 seconds for medium car flow and 3 seconds for dense car flow. Of course, one should also consider the realistic scenario of a certain intersection. For example, for an intersection with good cellular coverage, one should choose lower maximum delay value to achieve higher performance. 

In reality, a latency over 3 seconds is very rare. For example, if we model the latency as a Rayleigh distribution \cite{kwon2015tutorial} and consider an under-performed network with average latency as high as 500 ms (very under-performed), then the probability that the latency is more than 3 seconds is $6.1\times 10^{-13}$. This means that if one forwards information every 100ms, to eventually find one message having latency larger than 3 seconds, one need to wait $5196.8$ years.

Future research should consider introducing Recurrent Neural Network (RNN) architectures, e.g., Long Short-Term Memory (LSTM) as one of the DQN layers. The benefit of such a structure is that the topology of the network will remain the same for different maximum delay. This will enable curriculum learning for different delay which, in turn, will increase the stability of training (refer to section \ref{S:simulations}).

\section{Conclusion}
In this paper, a Deep Reinforcement Learning based delay-tolerant ITSC algorithm is proposed for intelligent traffic signal control. The algorithm is designed to mitigate the performance loss and eliminate the underlying safety issues in cloud-based ITSC systems caused by network latency. 

Simulation results clearly indicate that the algorithm will be able to tolerate up to 3 seconds of total delay without major performance loss. Therefore, the proposed algorithm is suitable for real world deployment, and hence, is of interest for infrastructure-based as well as infrastructure-free ITSC systems based on cloud computing as well as other technologies that requires a remote computation resource.
\bibliographystyle{IEEEtran}
\bibliography{ref}
\end{document}